\documentclass{icrc2009}

\usepackage{graphicx}   % for including figures
\usepackage[caption=false]{caption}    % for captions
\usepackage[font=footnotesize]{subfig} % subfig.sty for a double column floating figure using two subfigures
\usepackage{fixltx2e}
\usepackage{url}

\newcommand{\shorttitle}[1]%
{\markboth{Proceedings of the 31\MakeLowercase{$^{st}$} ICRC, {\L}\'{o}d\'{z} 2009}{#1} }
\newcommand{\etal}{\MakeLowercase{\textit{et al. }}} % "et al."

\begin{document}
\title{VERITAS observations of HESS J0632+057}

\author{\IEEEauthorblockN{Gernot Maier\IEEEauthorrefmark{1} for the VERITAS collaboration\IEEEauthorrefmark{2}}
			  
                            \\
\IEEEauthorblockA{\IEEEauthorrefmark{1}Department of Physics, McGill University, H3A 2T8 Montreal, QC, Canada (maierg@physics.mcgill.ca)}
\IEEEauthorblockA{\IEEEauthorrefmark{2} see R.A. Ong et al (these proceedings) or http://veritas.sao.arizona.edu/conferences/authors?icrc2009}}

% please write the preseter's name and short title (3-4 words maximum)
%    which will appear at the header of the even pages.
\shorttitle{G.Maier \etal VERITAS observations of HESS J0632+057}
\maketitle

\begin{abstract}

HESS J0632+057 is one of only two unidentified high energy gamma-ray sources which appear to be point-like in nature. It is possibly associated with the massive star MWC 148 and has been suggested to resemble known TeV binary systems like LS I +61 303 or LS 5039. These binaries are rare and extreme (only three TeV binaries are known to date), and their gamma-ray emission mechanism has not been understood.

HESS J0632+057 was observed by VERITAS, an array of four 12 m imaging atmospheric Cherenkov telescopes, in 2006, 2008 and 2009. Based on these observations we present evidence for variability in the high energy gamma-ray emission from HESS J0632+057. 
 \end{abstract}

 
\section{Introduction}

Point-like gamma-ray sources stand out among the many galactic VHE objects with spatially extended gamma-ray emission.
High-mass X-ray binaries constitute the only known class of galactic objects with variable point-like VHE emission; this class 
 currently contains three members only: PSR B1959-63/SS 2883 \cite{Aharonian-2005b}, LS 5039 \cite{Aharonian-2005c} and
LS I +61 303 \cite{Albert-2006, Acciari-2008}.

HESS J0632+057 is one of about $\sim20$ very high energy (VHE) gamma-ray sources with no known counterpart at other
wavelengths.
Gamma-ray emission was discovered by the High Energy Stereoscopic System (H.E.S.S.) 
during observations of the Monoceros Loop supernova remnant in 2004 and 2005 \cite{Aharonian-2007}.
HESS J0632+057 appears to be point-like within experimental resolution; the limit on the size of the emission region was given as 2' (95\%  confidence level).
The reported flux of gamma rays with energies above 1 TeV corresponds to about 3\% of the flux of the Crab
Nebula, with a differential photon spectrum  consistent with a power-law function with 
index of $2.53\pm0.26_{\mathrm{stat}}\pm0.20_{\mathrm{sys}}$.

Possible associations considered by \cite{Aharonian-2007} are the Monoceros Loop Supernova remnant, the weak X-ray source 1RXS J063258.3+054857,
the B0pe-star MWC 148 (HD 259440) and the unidentified GeV gamma-ray source 3EG J0634+0521 \cite{Hartman-1999}.
An association of HESS J0632+057 with the Monoceros loop SNR is unlikely given the point-like nature of the
gamma-ray emission and the absence of correlation of possible target material with the position of the VHE source 
\cite{Aharonian-2007}.
Follow-up X-ray observations with \emph{XMM-Newton} by \cite{Hinton-2009} revealed a variable X-ray source
(XMMU J063259.3+054801) with a position compatible with HESS J0632+057 and MWC 148.
3EG J0634+0521 is, as expected from the reported flux, not in the Fermi bright gamma-ray source list \cite{Abdo-2009}.

TeV binaries show variable emission of gamma rays, likely connected to changes in physical parameters 
associated with the orbital movement.
VHE gamma-ray production in these binaries is explained by the acceleration of charged particles in
accretion-powered relativistic jets \cite{Taylor-1984,Mirabel-1994}
or in shocks created by
the collision of the expanding pulsar wind with the wind from the stellar companion \cite{Maraschi-1981}.
Subsequent inverse-Compton scattering on low-energy stellar photons (leptonic models) or proton-proton collisions
(hadronic models)  produces gamma-rays at GeV and TeV energies.
While there has been no compact companion discovered for MWC 148, the point-like nature of the VHE
emission combined with the variable X-ray emission can easily be explained by a production 
scenario similar to those in TeV binaries.

An alternative scenario is that MWC 148 is a representative of a new type of VHE emitter as proposed by
\cite{Babel-1997} and \cite{Townsend-2007}.
In their picture strong magnetic fields around the massive star lead to magnetically channeled wind shocks in which
second-order Fermi acceleration might occur.
However, it is not clear if the circumstellar environment of MWC 148 is strongly magnetized,
or if this acceleration mechanism is able to produce particles of sufficiently high energy to produce a measurable TeV flux.

We present here results from observations of HESS J0632+057 with VERITAS. 
A detailed description of these observations and of contemporaneous X-ray observations with \emph{Swift} can be found
in \cite{Acciari-2009}.
 
 \section{Observations \& Results}
 
\begin{table*}[!h]
\caption{Details of the VERITAS and H.E.S.S. \cite{Aharonian-2007} observations of \mbox{HESS J0632+057}.
The VERITAS analysis results are given for $E>1$ TeV.
Upper limits $\Phi_{\gamma, \mathrm{UL}} (E>1$ TeV$)$
are given at 99\% confidence level 
(after \cite{Helene-1983}).
The integral fluxes and 1 $\sigma$ errors above 1 TeV reported by H.E.S.S. are listed for comparison \cite{Aharonian-2007}.
}
 \label{table:observations}
 \centering
 \begin{tabular}{|l|l|c|c|c|c|}
 \hline
 Name & Date range &  Elevation  & Obs. & significance & Flux or upper flux limit  \\
            &                        &     range      & time &   [$\sigma$]  &  [$10^{-13}$ cm$^{-2}$s$^{-1}$]  \\
            &                        &                      & [min] &                      &  \\
 \hline
 \hline
H.E.S.S. P1 & Dec 2004 &   &282 & & $6.3\pm1.8$  \\
H.E.S.S. P2 & Nov 2005 - Dec 2005 & &  372  & & $6.4\pm1.5$ \\
\hline
\hline
VERITAS & & & & & \\
\hline
Set I  & Dec 16 2006 - Jan 25 2007  & 55-65$^{\mathrm{o}}$  & 580 &  -0.9 & $<4.2$   \\
Set II  & Dec 30 2008 - Jan 03 2009  & 59-65$^{\mathrm{o}}$  & 560 & 1.3 & $<4.2$  \\
Set III  & Jan 26 2009 - Jan 30 2009  & 59-65$^{\mathrm{o}}$  & 722 & 1.0 & $<$ 3.6  \\
\hline
& total & & 1862 & 1.0 & $<$ 2.6 \\
   \hline
  \end{tabular}
  \end{table*}

VERITAS is an array of four imaging atmospheric-Cherenkov telescopes
located at the Fred Lawrence Whipple Observatory
in southern Arizona.
It combines
 good energy (15-20\%) and angular
($\approx 0.1^{\mathrm{o}}$) resolution
with
 a large effective
area (up to $10^5$ m$^2$) over a wide energy range (100 GeV
to 30 TeV).
The field of view of the VERITAS telescopes is 3.5$^{\mathrm{o}}$.
The high sensitivity of VERITAS enables
the detection of sources with a flux of 1\% of the Crab
Nebula in less than 50 hours of observations.
For more details on the VERITAS
instrument, see e.g.~\cite{Acciari-2008}.

\begin{figure*}[!t]
\centering
\includegraphics[width=3.0in]{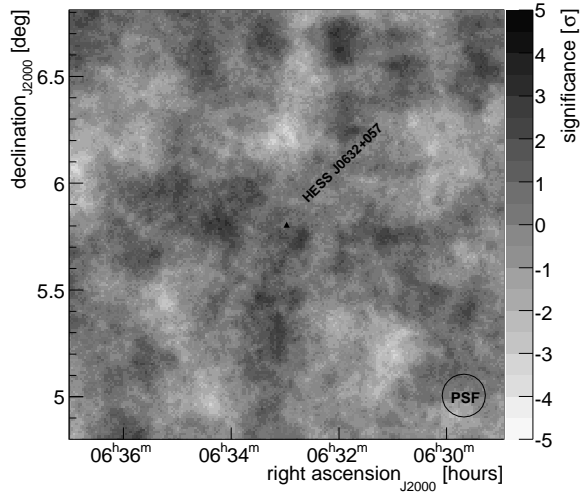}
\caption{
Significance map of the region around HESS J0632+057 for the whole VERITAS data set.
The location of HESS J0632+057 is indicated by a black triangle.
The background is estimated using
the "ring background" method.
The energy threshold is 720 GeV.
The circle at the bottom right indicates the angular resolution of the VERITAS observations.
}
\label{fig-skyPlot}
\end{figure*}

VERITAS observed the sky around HESS J0632+057 during three periods 
 in
December 2006, December 2008 and January 2009 (see Table \ref{table:observations} 
for details).
For each period, data equivalent to about 10 hours of observations passed data quality selection criteria,
which remove data taken during bad weather or with
hardware-related problems.
Data were taken on moonless nights 
in wobble mode, wherein
the source was positioned 0.5 degrees from the camera center with the offsets in
different positions for different runs.
The first data set (Set I) consists of observations taken
during the construction phase of VERITAS
with only 3 telescopes.
These observations were pointed towards the centre of the Monoceros region (at an angular distance of $\sim0.5^{\mathrm{o}}$ from
HESS J0632+057),
while observations in
the second and third set were targeted around the reported
position of HESS J0632+057.

The data analysis steps consist of image calibration and cleaning,
second-moment parameterization of these images \cite{Hillas-1985},
stereoscopic reconstruction of the event impact position and direction,
gamma-hadron separation, spectral energy reconstruction
(see e.g.~\cite{Krawczynski-2006})
as well as the generation of photon sky maps.
The majority of the far more numerous background events are rejected by comparing the shape of the event images in each 
telescope with the expected shapes of gamma-ray showers modeled by Monte Carlo simulations.
These \textit{mean-reduced-scaled width} and \textit{mean-reduced-scaled length}
cuts (see definition in \cite{Krawczynski-2006}), and an additional cut on the arrival direction of the incoming gamma ray ($\Theta^2$)
reject more than 99.9\% of the cosmic-ray background while keeping 45\% of the gamma rays.
The cuts applied here are:
integrated charge per image $>$ 1200 digital counts ($\approx$225 photoelectrons), 
 mean-reduced-scaled width/length between
-1.2 and 0.5,
 and \mbox{$\Theta^2 < 0.015$ deg$^2$}
(\mbox{$\Theta^2 < 0.025$ deg$^2$} for the 3-telescope data set).
The background in the source region is estimated from the same field of view
using the ``reflected-region'' model \cite{Aharonian-2001} and the ``ring-background'' model \cite{Aharonian-2005a}.
These cuts were chosen to provide good sensitivity at high energies, and to facilitate comparison
with the results reported in \cite{Aharonian-2007}.
The resulting energy threshold is 720 GeV.

\begin{figure*}[!t]
\centering
\includegraphics[width=3.0in]{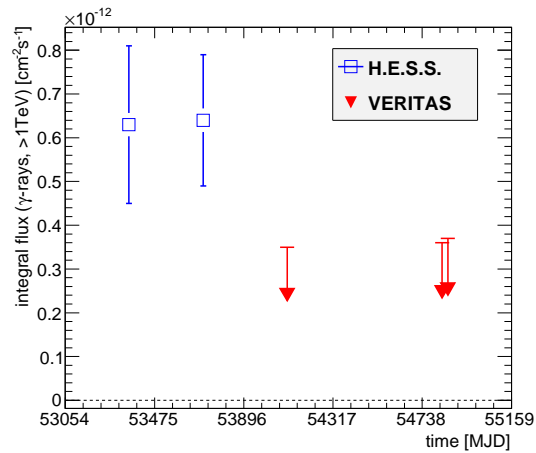}
\caption{
The light curve above 1 TeV from HESS J0632+057 is shown assuming a spectral
shape of $dN/dE \propto E^{-\Gamma}$ with $\Gamma=2.5$. 
The downwards pointing arrows show the 99\% confidence limits derived here from the VERITAS data.
The H.E.S.S. fluxes are taken from \cite{Aharonian-2007}.
}
\label{fig-lightcurve}
\end{figure*}

Results for each of the three VERITAS data sets, as well as for the total
observation can be found in Figures \ref{fig-skyPlot}, and \ref{fig-lightcurve},  and   Table \ref{table:observations}.
Figure \ref{fig-skyPlot}  shows a sky map  of the significance at energies above 720 GeV observed in the region
around HESS J0632+057.
The distribution of significances in the sky map is consistent with the expected distribution from
a field with
no gamma-ray source present. The significance at the position of HESS J0632+057 is $2.1\sigma$ ($1\sigma$ for
an energy threshold of 1 TeV, see Table \ref{table:observations}).
There is no significant evidence for $\gamma$-ray like events from HESS J0632+057 during the 31 hours of observations
with VERITAS.
The flux upper limit (E$>$1 TeV) for the complete data set at the 99\% confidence level \cite{Helene-1983}
assuming a power-law like source spectrum with a spectral index of $\Gamma=2.5$ is
 $2.6\times 10^{-13}$ cm$^{-2}$s$^{-1}$ (about 1.1\% of the flux of the
Crab Nebula; see Table \ref{table:observations}).
This flux limit is $\sim2.4$ times lower than the flux reported by H.E.S.S in \cite{Aharonian-2007};
see Figure \ref{fig-lightcurve} for a light-curve.
The probability for a non-variable flux of high-energy gamma rays from HESS J0632+057 is 
derived from the VERITAS data
 and the average of the reported fluxes from H.E.S.S. using a $\chi^2$-test. 
%The average H.E.S.S. flux is ($6.4\pm1.15)\times 10^{-13}$ cm$^{-2}$s$^{-1}$. 
The test gives a $\chi^2$ of 15.8 with 1 degree of freedom, corresponding to a probability of 0.007\% (about 4$\sigma$).

\section{Conclusions}

The non-detection of HESS J0632+057 by VERITAS provides evidence for variability in the flux of gamma-rays with
energies above 1 TeV. 
Variability has also been found in X-rays \cite{Hinton-2009} \cite{Acciari-2009} \cite{Falcone-2009} and radio \cite{Skilton-2009}.

The VHE emission and variability can be easily explained if MWC 148 is part of a binary system and the high-energy photons
are produced in a similar way to those in LS I +61 303 or LS 5039.
Particle acceleration and VHE-emission from massive stars with strong magnetic fields has also been suggested.
A confirmation that MWC 148 is surrounded by sufficiently strong magnetic fields, along with further theoretical work
to explain the variability in the gamma-ray emission, is needed to establish this potential new class of
galactic gamma-ray sources.

Future multiwavelength observations, combined with results from ground-based and space-based gamma-ray observatories
will provide a deeper understanding of the true nature of HESS J0634+057.

\section{Acknowledgments}
This research is supported by grants from the US Department
of Energy, the US National Science Foundation, and the Smithsonian
Institution, by NSERC in Canada, by Science Foundation
Ireland, and by STFC in the UK.
We acknowledge the
excellent work of the technical support staff at the FLWO and
the collaborating institutions in the construction and operation
of the instrument.

\end{document}